\documentclass[nanomaterials,aricle,submit,pdftex,moreauthors]{Definitions/mdpi}
\usepackage{soul} 
\usepackage{color}
\definecolor{vs}{rgb}{0.1,0.4,0.1}                  

\newcommand{\del}[1]{}                             
\firstpage{1}
\pubvolume{1}
\issuenum{1}
\articlenumber{0}
\pubyear{2022}
\copyrightyear{2022}
\datereceived{6 May 2022}
\dateaccepted{27 May 2022}
\datepublished{}
\hreflink{https://doi.org/} 

\Title{Nature of the Poynting Vector Field Singularities in Resonant Light Scattering by Nanoparticles}

\TitleCitation{Nature of the Poynting vector field singularities...}

\Author{Michael I. Tribelsky $^{1,2}$\orcidA{}, Boris Y. Rubinstein$^{3}$}

\AuthorNames{Michael I. Tribelsky and Boris Y. Rubinstein}

\AuthorCitation{Tribelsky, M.; Rubinstein, B.}

\address{%
$^{1}$ 
M. V. Lomonosov Moscow State University, Moscow, 119991, Russia; mitribel\_at\_gmail\_dot\_com\\
$^{2}$ 
National Research Nuclear University MEPhI (Moscow Engineering Physics Institute), Moscow, 115409, Russia\\
$^{3}$ 
Stowers Institute for Medical Research, 1000 E. 50th St., Kansas City, MO 64110, USA}

\abstract{Singularities of the Poynting vector field subwavelength patterns in resonant light scattering by nanoparticles are discussed and classified. There are two generic types of the singularities, namely, (i) the singularities related to the vanishing of the magnetic (and/or electric) field at the singular points and (ii) the singularities related to the formation of standing waves in proximity to the singular points. The connection of these types of singularities to the topology of the singular points, space dimension (3D vs. 2D), and energy conservation law are revealed. In particular, it is shown that in 2D cases in non-dissipative media, the energy conservation reduces the possible types of generic singular points to saddles and centers only. In 3D cases, a universal expression connecting different components of the Poynting vector and valid for any generic singularities is derived and numerically checked for various types of singular points.}


\begin{document}

\section{Introduction}
{In addition to the purely academic interest, the structure of the electromagnetic field in the immediate vicinity of a nanoparticle irradiated by a laser beam is of utmost importance for numerous nanomaterial science problems, see, e.g., \cite{AlZoubi2022,AlZoubi2022a,Chen2022,Zheng2022,Li2021,Xiao2017,zhang2015} and references therein.}
It is known that the topological structure of the Poynting vector field in resonant light scattering by nanoparticles may be complicated. Specifically, the near-field wave zone may include singular points of different types~\cite{Luk2004:PRB,Zheludev2005:OE,OptJ_2006,Tribelsky:2006jf}. The eigenvalues (roots of the characteristic equations) of a singular point completely determine the topological structure of the Poynting vector field in the proximity of this point~\cite{Novitsky2009}, while the spatial position of given singular points determines the global topological structure of the field as a whole. For these reasons, singularities are essential in understanding any vector field's structure.
This is especially true in the case of the Poynting vector, since the corresponding field describes the energy flow, while the direct experimental measurement of this flow at the nanoscale resolution is a very challenging task.

{At present, studies of singularities in electromagnetic fields constitute a separate discipline called {\it singular optics}. The results of these studies are presented in a vast collection of research papers, reviews, book chapters, and monographs; see, e.g., \cite{Angelsky2021,Yue2019,Gao2014,Dennis2009,Novitsky2009,Mokhun2007}.}
{A detailed Poincar\'{e}-type classification of the Poynting vector field's singular points was made by Novitsky and Barkovsky~\cite{Novitsky2009}. Authors of further publications were basically focused on the study of singularities in more complicated cases, such as, for example, in a non-diffractive tractor beam~\cite{Gao2014}.}

{Meanwhile, some fundamental questions related to peculiarities of singular points in subwavelength patterns at the resonant scattering of a monochromatic plane linearly polarized electromagnetic wave by simple objects, such as a spatially uniform nanosphere or right circular cylinder, remain open. The goal of this paper is to answer several of these questions. Specifically, we inspect how, in these cases, the singularities of the Poynting vector field are related to the topological structure of the $\mathbf{E}$ and $\mathbf{H}$ fields, dimension of space, energy circulation, and energy conservation law.}

%
\section{Methods}
We discuss the scattering of a plane linearly polarized monochromatic electromagnetic wave with the temporal dependence of the fields $\sim \exp(-i\omega t)$ by a homogeneous sphere of radius $R$ or a right infinite circular cylinder with the base radius $R$. The scattering particles are non-magnetic, so their permeability $\mu = 1$. {We employ an analytical study of the vector Poynting field lines (streamlines) in the vicinity of a generic singularity supplemented by the energy conservation law. General arguments are illustrated by a detailed discussion of several specific cases. Their analysis is based on the exact solutions to the related problems; see, e.g.,~\cite{Bohren::1998}. The symbolic calculations and visualization of the results are made with the help of the Wolfram {\it Mathematica} software.}

\section{Results and Discussion}
\subsection{Sphere}
{For the time being, we use the term {\it singularity} in its {\it geometric} meaning only, i.e., applying it to a point where the Poynting vector streamlines asymptotically merge or intersect.} To begin with, we discuss the scattering by a subwavelength sphere. {Following conventional notations~\cite{Bohren::1998}, we suppose that the sphere's center is the origin of the coordinate frame; the incident wave propagates along the positive direction of the $z$-axis, and its vector $\mathbf{E}$ oscillates along the $x$-axis.}
The symmetry of the problem requires the plane $xz$ to be invariant for the streamlines. Moreover, the calculations show that, for the problem in question, all singular points belong to this plane~\cite{Luk2004:PRB,Zheludev2005:OE,OptJ_2006,Tribelsky:2006jf}. Then, it seems that the analysis of singularities in the plane is sufficient. In this case, we have only four generic types of singularities: a saddle, node, focus, and center.

Next, for a saddle, the streamlines corresponding to the stable and unstable manifolds intersect at the singular point. In the case of a focus or node, all streamlines asymptotically merge in (for stable) or emerge from (for unstable) the singularity. Since, by definition, the Poynting vector is tangential to its streamlines, the intersection or merging of the streamlines means that the direction of the Poynting vector in a saddle, node, or focus is not uniquely defined. The same conclusion is valid for a center-type singular point: in its proximity, the streamlines are closed ellipses whose diameters vanish as one approaches the singularity.

An undefinable vector direction at a certain point is consistent with the nature of a vector field if, and only if, the vector vanishes at this point. Thus, the Poynting vector must vanish at the singular points. Then, seemingly, in the most interesting cases of foci (optical vortices) or nodes, we encounter a violation of the energy conservation law. Indeed, by definition, streamlines indicate the direction of the energy flow. Therefore, in the case of a stable focus or node, electromagnetic energy must accumulate at the singular point. This accumulation either is inconsistent with any fixed pattern of the scattered field, or there should be a sink for the energy flow directed to the singularity. The flow is inverted for an unstable singularity. Therefore, instead of a sink, a source is required in this case.

{We readily resolve the paradox by recalling that our space is three-dimensional (3D). The employed problem formulation (a scattering particle embedded in an infinite ambient medium or vacuum; the incident wave comes from infinity, and the scattered radiation goes to infinity) is physically meaningful, provided the ambient medium is non-dissipative. In such a medium, the divergence of the Poynting vector must identically equal zero. ~\cite{landau2013electrodynamics}.}

{Suppose we have a 2D stable singularity in the invariant plane outside the scattering sphere. Then, the vanishing divergence requires that in 3D, there must be an unstable manifold in the transversal direction. The unstable focus (node) must have a stable manifold in a transversal direction.}

{To prove the above reasoning quantitatively and to show that, in addition to the discussed geometric meaning, the condition $\mathbf{S}=0$ determines the Poincar\'{e}-type singularities in a phase space of equations describing the streamlines, we have to explicitly obtain these equations. To this end, it is convenient to present the streamlines in the following parametric form: $\mathbf{r}=\mathbf{r}(t)$, where $t$ plays the role of effective time~\cite{Novitsky2009}. Then, the ``velocity,'' $d\mathbf{r}/dt$ should be tangential to the ``trajectory'' (streamline), i.e., parallel to the direction of $\mathbf{S}(\mathbf{r}(t))$ at the same point of the streamline. Thus, the equation governing the ``dynamics'' of the streamlines should be as follows:
\begin{equation}\label{eq:dr/dt}
  \frac{S_0}{c}\frac{d\mathbf{r}}{dt} = \mathbf{S}(\mathbf{r}),
\end{equation}
where $S_0$ and $c$ are arbitrary constants with the proper dimensions. In what follows, we suppose them to be equal to the intensity of the incident wave and the speed of light, respectively.}

{Eq.~\eqref{eq:dr/dt}, written in components, splits into three coupled equations for $x,y$, and $z$. The conditions determining their nullclines are as follows: $S_{x,y,z}=0$. At singular points of Eq.~\eqref{eq:dr/dt} all three conditions $S_{x,y,z}=0$ should simultaneously hold, i.e., the Poynting vector should vanish. That is to say, the geometric and Poincar\'{e} criteria for singular points give rise to identical results.}

{In further analysis, it is convenient to introduce dimensionless variables. Conventionally, the natural spatial scale for the problem is the wavelength of the incident radiation. Accordingly, the corresponding coordinate scale transformation would be $\mathbf{r}_{new}= k\mathbf{r}$, where $k = \omega/c$ is the wavenumber of the incident wave. However, here we are interested in subwavelength patterns. Then, in our case, it is more convenient to normalize the spatial variables on the radius of the scattering sphere, $R$, i.e., to perform the scale transformation $\mathbf{r}_{new} = \mathbf{r}/R$. Similarly, $t_{new}=tc/R$ and $\mathbf{S}_{new}=\mathbf{S}/S_0$. Since below, only the dimensionless quantities will be employed, the index {\it new} will be dropped.}

Bearing in mind that the plane $y=0$ is invariant, i.e., in this entire plane $S_y=0$; shifting the origin of the coordinate frame to the singularity, and expanding $\mathbf{S}(\mathbf{r})$ in powers of small $\mathbf{r}$; we obtain that, in the vicinity of the singularity, the streamlines are defined by the following equations:
\begin{eqnarray}
  \frac{dx}{dt} &=& S_x(x,y,z) \approx s_{xx}x + s_{xy}y + s_{xz}z, \label{eq:dx/dt} \\
  \frac{dy}{dt} &=& S_y(x,y,z) \approx s_{yy}y, \label{eq:dy/dt}\\
  \frac{dz}{dt} &=& S_z(x,y,z) \approx s_{zx}x + s_{zy}y + s_{zz}z, \label{eq:dz/dt}
\end{eqnarray}
where $s_{nm}=\left(\frac{\partial S_n}{\partial x_m}\right)_{\! sin}$, and subscript {\it sin} means that the derivatives are taken at the singular point. Here, $S_n$ and $x_m$ stand for any of the three components of vectors {\bf S} and {\bf r}, respectively. Note that the condition div$\,\mathbf{S}=0$ imposes the following connections between \mbox{coefficients $s_{nm}$:}
\begin{equation}\label{s1+s2+s3}
  s_{xx}+s_{yy}+s_{zz}=0.
\end{equation}

{Looking for a solution to Eqs.~\eqref{eq:dx/dt}--\eqref{eq:dz/dt} in the form $x_m = const_m e^{\kappa t}$ (where, as before, $x_m$ stands for any coordinate $x,y,z$), and equalizing the determinant of the resulting linear algebraic equation to zero, we obtain a cubic characteristic equation. The latter splits into a detached equation $\kappa = s_{yy}$, following from Eq.~\eqref{eq:dy/dt}, and a quadratic one, following from Eqs.~\eqref{eq:dx/dt},~\eqref{eq:dz/dt}. Finally, this procedure gives rise to
\begin{equation}\label{eq:3D_eigenval}
  \kappa_1 = -2\gamma; \;\; \kappa_{2,3} = \gamma \pm \alpha,
\end{equation}
where we have introduced the following notations:
\begin{equation}\label{eq:3D_alpha_gamma}
 \alpha = \frac{\sqrt{(s_{xx}-s_{zz})^2+4 s_{xz} s_{zx}}}{2}\;;\;\; \gamma = \frac{(s_{xx} + s_{zz})}{2} \equiv - \frac{s_{yy}}{2},
\end{equation}
see also Eq.~\eqref{s1+s2+s3}.}

Eqs.~\eqref{eq:dx/dt}--\eqref{eq:dz/dt} are exactly integrable, but the corresponding expressions are cumbersome. For this reason, below, only the solution describing the streamlines in the invariant plane $y=0$ is presented. It has the following form:
\begin{eqnarray}
  x(t) &=& e^{\gamma  t}\left[\frac{2 c_2 s_{xz}+c_1 (s_{xx}-s_{zz})}{2\alpha }\sinh {\alpha  t}+c_1 \cosh {\alpha  t}\right], \label{eq:x(t)} \\
  z(t) &=& e^{\gamma  t} \left[\frac{2 c_1 s_{zx}+ c_2 (s_{zz}-s_{xx})}{2 \alpha }\sinh{\alpha  t}+c_2 \cosh {\alpha  t}\right],  \label{eq:z(t)}
\end{eqnarray}
where $c_{1,2}$ are arbitrary constants of integration. According to the implemented normalization procedure, all quantities in Eqs.~\eqref{eq:x(t)}, \eqref{eq:z(t)} are dimensionless.

{At $(s_{xx}-s_{zz})^2+4 s_{xz} s_{zx}>0$ all $\kappa_{1,2,3}$ are real. In the invariant plane, this case corresponds either to a saddle ($|\gamma|<|\alpha|$) or a node ($|\gamma|>|\alpha|$). For the latter, the signs of $\kappa_1$ and $\kappa_{2,3}$ are always opposite; see Eq.~\eqref{eq:3D_eigenval}. This means that a stable 2D node ($\kappa_{2,3} <0$) is unstable in $y$-direction and vice versa, in full agreement with the above qualitative reasoning.}

{At  $(s_{xx}-s_{zz})^2+4 s_{xz} s_{zx}<0$ eigenvalues $\kappa_{2,3}$ are complex, i.e., in the invariant plane, the singularity is a focus. It is stable at $\gamma<0$ and unstable at the opposite sign of $\gamma$.}

{Note that $y(t)=const \cdot\exp(-2\gamma t)$; see Eqs.~\eqref{eq:dy/dt},~\eqref{eq:3D_eigenval},~\eqref{eq:3D_alpha_gamma}. Thus, as pointed out above, the singularity, which, in the invariant plane, is a focus, in 3D is a focus-saddle. Moreover, a stable 2D focus has an unstable manifold in the $y$-direction, while an unstable focus has a stable $y$-directed manifold. In both cases, the index in $y$-direction ($-2\gamma$) has the opposite sign and twice as much modulus as that for the expansion (contraction) index for the streamlines in the invariant plane ($\gamma$).}

\begin{figure}[H]
\includegraphics[width=14.3 cm]{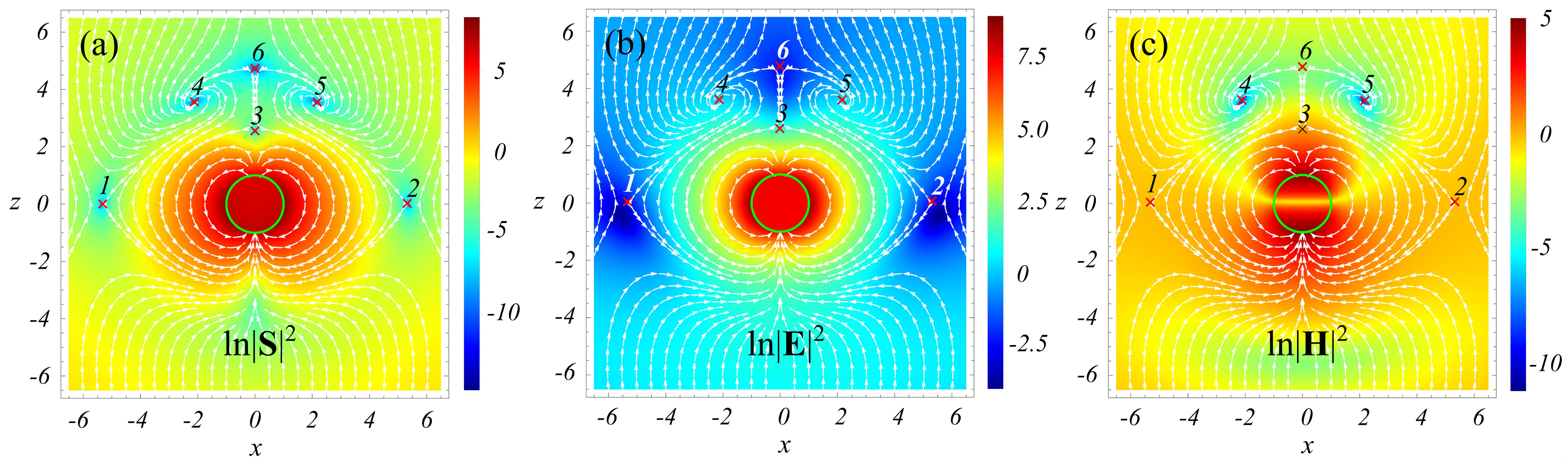}
\caption{{Light scattering by a sphere in a vacuum. Streamlines for the Poynting vector field and the logarithm of the square of the modulus of the Poynting vector (a), electric (b), and magnetic (c) fields in the invariant plane passing through the sphere's center and coinciding with the plane of vector $\mathbf{E}$ oscillations. $q=0.3$, $\varepsilon=-2.17$. The pattern is symmetric with respect to the line $x=0$. Crosses marked with numbers are the positions of singular points. Points 4 and 5 are foci; the rest are saddles. $|\mathbf{S}|^2=0$ at all singular points. The foci and saddles are the {\it H-field-induced} and {\it polarization-induced} singularities, respectively. See text for details.}\label{fig:Sphere}}
\end{figure}
To illustrate the above results, the streamlines of the Poynting vector superimposed with the density plots of $\ln |\mathbf{S}|^2$, $\ln |\mathbf{E}|^2$, and  $\ln |\mathbf{H}|^2$ are presented in Fig.~\ref{fig:Sphere}. The plots correspond to the following values of the problem parameters: $q=0.3;\;\; \varepsilon = -2.17$. Here, the size parameter \mbox{$q = kR = 2\pi R/\lambda$;} and $\lambda$ stands for the wavelength of the incident wave. The selected value of $\varepsilon$ lies close to the dipole resonance point {(at $q=0.3$, the maximum of the dipole resonance line is situated at \mbox{$\varepsilon \approx -2.22$}) and approximately corresponds to aluminum at $\lambda \approx 142$~nm~\cite{Polyanskiy}. The color bars in Fig.~\ref{fig:Sphere} designate the values of $\ln |\mathbf{S}|^2$,  $\ln |\mathbf{E}|^2$, and  $\ln |\mathbf{H}|^2$, where the fields are normalized on the corresponding quantities in the incident wave.} The surface of the scattering sphere is shown as a green circle. Considerable enhancement of $|\mathbf{S}|^2$,  $|\mathbf{E}|^2$, and  $|\mathbf{H}|^2$ in the vicinity of the particle is explained by the accumulation of radiation from large downstream area and its transport to the particle (funnel effect~\cite{bohren1983can,tribelsky2022resonant}).

We recall that the spatial scale in the figures is normalized on $R$. Therefore, since $\lambda =2\pi R/q \approx 20.9 R$, at the given value of the size parameter, the patterns presented in the figure are essentially subwavelength. These patterns indicate that:
\begin{description}
  \item[(i)] $|\mathbf{S}|$ does vanish at all singular points;
  \item[(ii)] $\mathbf{S}$ vanishes at the foci owing to vector $\mathbf{H}$ vanishing;
  \item[(iii)] in contrast, at the saddle-type singular points neither $\mathbf{E}$ nor $\mathbf{H}$ turn to zero.
\end{description}
{Case (iii) is explained by forming a standing wave in the vicinity of the saddle-type singularities. Indeed, a saddle has a stable manifold. This manifold attracts the Poynting vector streamlines so that, in proximity to the singular point, they are directed toward the singularity, i.e., {\it in strictly opposite directions at the opposite sides of the latter}. In other words, two counter-propagating traveling waves are formed. The problem's symmetry requires the equality of the amplitudes of these waves at the singularity. The same arguments are valid for the unstable manifold. Then, though the amplitudes of each of the counter-propagating waves at the singularity remain finite, the Poynting vector vanishes, which agrees with Fig.~\ref{fig:Sphere}(a). Importantly, the scale of the region with the counter-propagating energy flows, in this case, is {\it much smaller} than the wavelength of the incident radiation. Following the notations of Ref.~\cite{Novitsky2009} we will call the (ii)- and (iii)-type singularities as {\it field-induce} and {\it polarization-induced}, respectively.}

The above arguments make it possible to estimate the in-plane and transversal components of the Poynting vector in the vicinity of a singular point. To this end, we shift the origin of the coordinate frame to the singular point in question and embed it in a right circular cylinder whose base is parallel to the invariant plane so that the singularity lies in the middle of the height of the cylinder. Let small $2|y|$ and $r$ be the height of the cylinder and its base radius, respectively. The total flux of the Poynting vector through the surface of the cylinder must be zero. This gives rise to the following equality:
\begin{equation}\label{eq:flux}
  2\pi r^2 |\langle S_y \rangle| = 4\pi r |y \langle S_r\rangle|,
\end{equation}
where $\langle S_{y,r}\rangle$ designate the mean values of the corresponding components of $\mathbf{S}$ at its bases ($\langle S_{y}\rangle$) and side wall ($\langle S_{r}\rangle$).

Bearing in mind that, according to Eq.~\eqref{eq:dy/dt}, $\langle S_y \rangle \approx s_{yy}y$, Eq.~\eqref{eq:flux} results in the following estimate connecting the mean radial components of the Poynting vector with the $y$-derivative of its $y$-component:

\begin{equation}\label{eq:S_y_vs_S_r}
 \left|\frac{\langle S_r \rangle }{ s_{yy}}\right|   \approx \frac{r}{2}.
\end{equation}
Note that the r.h.s of Eq.~\eqref{eq:S_y_vs_S_r} is universal and does not depend on the type of singularity.

We numerically checked the validity of expression \eqref{eq:S_y_vs_S_r} for all singularities presented in Fig.~\ref{fig:Sphere}, supposing  $y=0.1$ and $r$ varying from 0.001 to 0.2. The calculations show that, at the same values of $r$, the numerical values of $\langle S_r \rangle$ and $s_{yy}$ for the foci and saddles are quite different; however, for all these singularities, condition~\eqref{eq:S_y_vs_S_r} holds with high precision.

\begin{figure}[H]
\includegraphics[width=10.5 cm]{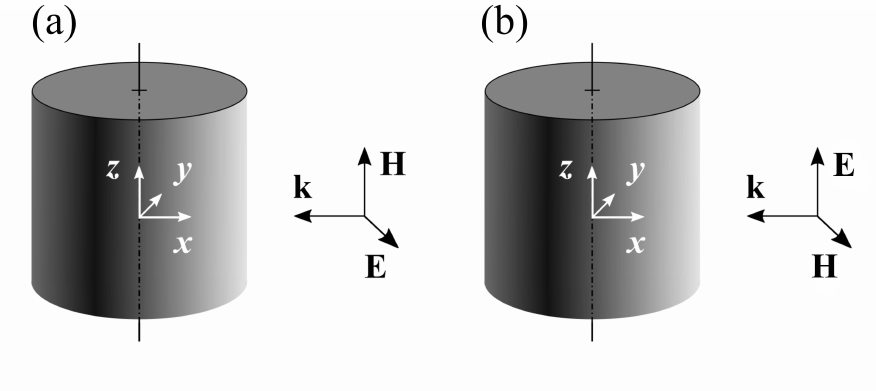}
\centering
\caption{{Mutual orientations of the cylinder and {\bf k}, {\bf E}, {\bf H} vectors of the incident wave. \mbox{TE polarization (a),} TM polarization (b).}\label{fig:TE_TM}}
\end{figure}

{\subsection{Cylinder}

In the case of light scattering by an infinite cylinder, the field pattern is two-dimensional (2D). Let us discuss how the dimension of the pattern affects singular points. Following the conventional problem formulation~\cite{Bohren::1998} we chose the coordinate frame with the $z$-axis coinciding with the cylinder's axis and wave vector of the incident wave antiparallel to the $x$-axis; see Fig.~\ref{fig:TE_TM}.

Then, under the assumptions mentioned above, instead of Eqs.~\eqref{eq:dx/dt}--\eqref{eq:dz/dt} we obtain the following set of equations, describing 2D streamlines:
\begin{eqnarray}
  \frac{dx}{dt} &=& S_x(x,y,z) \approx s_{xx}x + s_{xy}y, \label{eq:dx/dt_2D} \\
  \frac{dy}{dt} &=& S_y(x,y,z) \approx s_{yx}x+s_{yy}y. \label{eq:dy/dt_2D}
\end{eqnarray}}

{To begin with, we discuss the singularities outside the cylinder. In this case, the condition div$\,\mathbf{S}=0$ requires that
\begin{equation}\label{eq:div2D=0}
   s_{xx}+s_{yy}=0.
\end{equation}
We note that all quantities in Eqs.~\eqref{eq:dx/dt_2D}--\eqref{eq:div2D=0} are dimensionless.}

{Bearing in mind Eq.~\eqref{eq:div2D=0}, we obtain the set of the eigenvalues of Eqs.~\eqref{eq:dx/dt_2D}--\eqref{eq:dy/dt_2D}
\begin{equation}\label{eq:2D_indices}
  \kappa_{1,2} = \pm \sqrt{s_{xx}^2+s_{xy}s_{yx}}\;.
\end{equation}
Thus, the only possible types of non-degenerate singular points in this case are saddles at $s_{xx}^2>s_{xy}s_{yx}$, or centers at $s_{xx}^2<s_{xy}s_{yx}$. The equations describing the corresponding streamlines are as follows:
\begin{eqnarray}
  & & x(t) = c_1\cosh(\kappa t)+\frac{c_1s_{xx}+c_2s_{xy}}{\kappa}\sinh(\kappa t), \label{eq:x(t)_2D}\\
  & & y(t) = c_2\cosh(\kappa t)+\frac{c_1s_{yx}-c_2s_{xx}}{\kappa}\sinh(\kappa t), \label{y(t)_2D}
\end{eqnarray}
where $c_{1,2}$ are constants of integration. Since the r.h.s.' of Eqs.~\eqref{eq:x(t)_2D}--\eqref{y(t)_2D} are even functions of $\kappa$, these equations are invariant to the specific choice of the sign of $\kappa$ in Eq.~\eqref{eq:2D_indices}.}

Regarding singularities inside the cylinder, if the latter is made of a non-dissipative material, the case is the same as that discussed above. However, closed loops of the streamlines are forbidden if the imaginary part of the cylinder permittivity is not equal to zero, and hence the scattering is accompanied by dissipation. Then, the center-type singularities in the cylinder must transform into foci. This transformation was observed in numerical simulations~\cite{OptJ_2006}.

{Next, it is instructive to consider how the {\it field-induced} and {\it polarization-induced} 2D singularities look. To this end, we study the resonant light scattering using a cylinder with the size parameter $q=0.1$ and $\varepsilon = -1$ for the normal incidence of the TE-polarized light (the vector {\bf E} of the incident wave is perpendicular to the axis of the cylinder), as well as with $q=0.3,\;\varepsilon = 16$ for the TM-polarized light (the vector {\bf H} of the incident wave is perpendicular to the axis of the cylinder); see Fig.~\ref{fig:TE_TM}. The selected values of permittivity approximately correspond to aluminum at $\lambda \approx 113$~nm ($\varepsilon = -1$) and silicon at $\lambda \approx 565$~nm ($\varepsilon = 16$)~\cite{Polyanskiy}.}

\begin{figure}[H]
\includegraphics[width=14.3 cm]{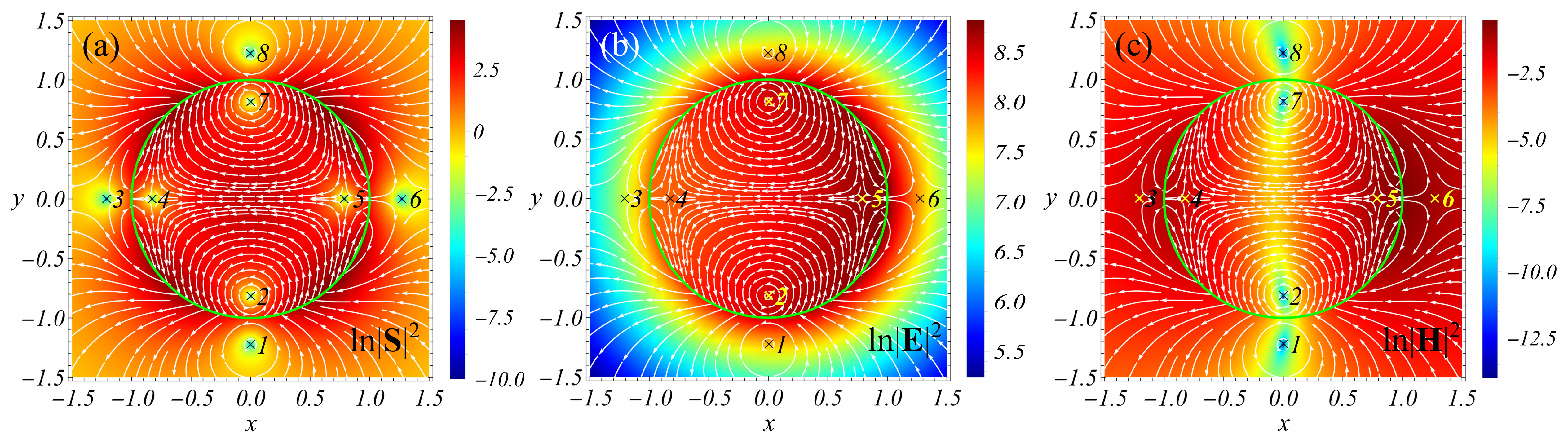}
\caption{{Light scattering by a cylinder in a vacuum. Streamlines for the Poynting vector field and the logarithm of the square of the modulus of the Poynting vector (a), electric (b), and magnetic (c) fields in the $xy$-plane. The TE polarization (vector {\bf E} of the incident wave lies in the plane of the figure). $q=0.1$, $\varepsilon=-1$. The pattern is symmetric with respect to the line $y=0$. A green circle designates the surface of the cylinder. Crosses marked with numbers are the positions of singular points. Points 1,2,7,8 are centers; the rest are saddles. $|\mathbf{S}|^2=0$ at all singular points. The centers and saddles are the {\it H-field-induced} and {\it polarization-induced} singularities, respectively. See text for details.}\label{fig:Cylinder_TE}}
\end{figure}

\begin{figure}[H]
\includegraphics[width=14.3 cm]{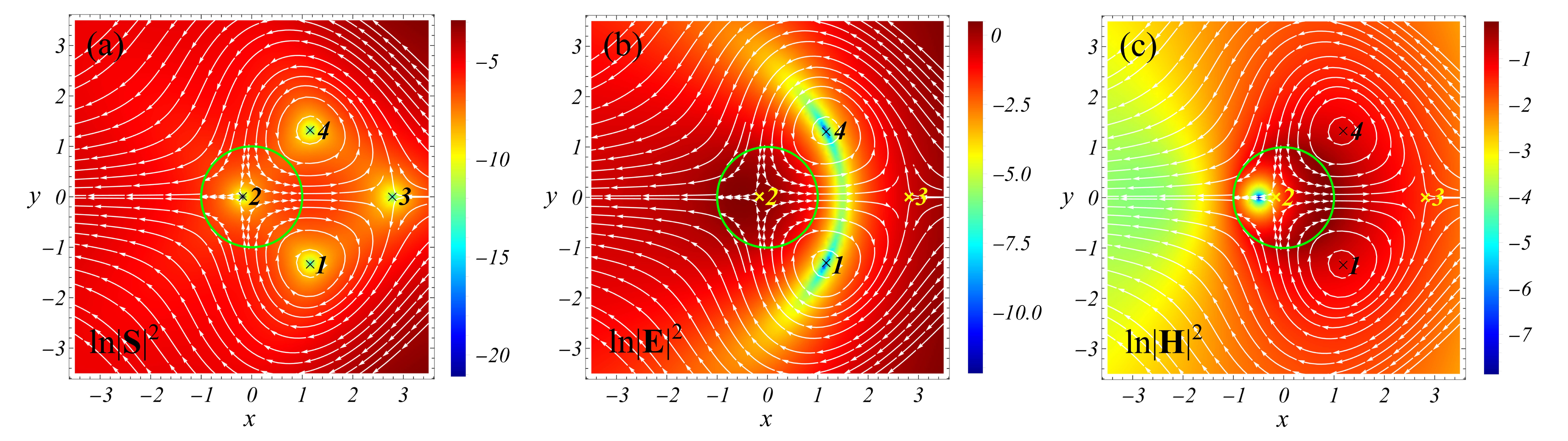}
\caption{{The same as that in Fig.~\ref{fig:Cylinder_TE} for the TM polarization (vector {\bf H} of the incident wave lies in the plane of the figure) and $q=0.3$, $\varepsilon=16$. Points 1 and 4 are centers; the rest are saddles. $|\mathbf{S}|^2=0$ at all singular points. The centers and saddles are the {\it E-field-induced} and {\it polarization-induced singularities}, respectively. See text for details.}\label{fig:Cylinder_TM}}
\end{figure}

{The results of the study are presented in Figs.~\ref{fig:Cylinder_TE}, \ref{fig:Cylinder_TM}. We observe that, in agreement with the 3D case, the saddles are {\it polarization-induced} singularities, while the centers are {\it field-induced}. Note that, if a {\it field-induced} singularity belongs to a pattern lying in the plane of vector {\bf E} oscillations, it appears to be \mbox{{\it H-field-induced}}. In contrast, if the pattern lies in the plane of vector {\bf H} oscillations, the singularity is \mbox{{\it E-field-induced}.}}
\unskip

\section{Conclusions}
Summarizing the results obtained, we can say the following:
\begin{itemize}
  \item {The geometric and Poincar\'{e} criteria for singularities give rise to the same condition: $\mathbf{S}(\mathbf{r})=0$ at singular points.}
  {\item In 3D problems optical vortices are {\it field-induced} singularities, while saddles are {\it polarization-induced}.
  \item In the vicinity of any 3D generic singularity, regardless of its specific type, the universal condition \eqref{eq:S_y_vs_S_r} must hold.
  \item If a {\it field-induced} singularity lies in an invariant plane, which is parallel to the plane of oscillations of vector {\bf E}, the singularity is {\it H-field-induced}. If the invariant plane is parallel to the plane of vector {\bf H} oscillations, the singularity is {\it E-field-induced}.
  \item In 2D problems, the condition div$\,\mathbf{S}=0$ reduces the types of possible non-degenerate singularities only to saddles and centers.}
  \item {In 3D problems, optical vortices in non-dissipative media have a focus-saddle structure. Importantly, for a stable focus, the saddle is unstable and vice versa. Moreover, the modulus of the contraction (expansion) index in the direction transversal to the plane of the focus is twice as much as that for the in-plane streamlines.}
 \end{itemize}
It also should be stressed that the characteristic spatial scale of all discussed patterns is much smaller than the incident radiation wavelength. We hope the unveiled peculiarities of the singular points shed new light on the vital problem of resonant light scattering by subwavelength particles and can help to enrich and optimize its practical applications.

\vspace{6pt}

%

\authorcontributions{
Conceptualization, M.T.; methodology, M.T; software, B.R.; 
investigation, M.T. and B.R.; 
writing---original draft preparation, M.T.; writing---review and editing, M.T. and B.R.; visualization, M.T. and B.R.; supervision, M.T.; 
funding acquisition, M.T. All authors have read and agreed to the published version of the manuscript.}

\funding{This research was funded by the Russian Foundation for Basic Research (Projects No. 20-02-00086) for the analytical study and the Russian Science Foundation (Project No. 21-12-00151) for the computer calculations.}

\conflictsofinterest{
The authors declare no conflict of interest.}



\abbreviations{Abbreviations}{
The following abbreviations are used in this manuscript:\\

\noindent
\begin{tabular}{@{}ll}
2D & two-dimensional\\
3D & three-dimensional\\
MDPI & Multidisciplinary Digital Publishing Institute\\
r.h.s. & right-hand side\\
TE &  Transverse Electric\\
TM &  Transverse Magnetic
\end{tabular}}

%
%
%

\reftitle{References}


\bibliography{Singularities_rev2}

\begin{thebibliography}{999}

\bibitem[Al~Zoubi \em{et~al.}(2022)Al~Zoubi, Allaf, Assfour, and
  Ko]{AlZoubi2022}
Al~Zoubi, W.; Allaf, A.W.; Assfour, B.; Ko, Y.G.
\newblock Concurrent Oxidation-Reduction Reactions in a Single System Using a
  Low-Plasma Phenomenon: Excellent Catalytic Performance and Stability in the
  Hydrogenation Reaction.
\newblock {\em ACS Appl. Mater. Interfaces} {\bf 2022}, {\em 14},~6740--6753.
\newblock
  doi:{\changeurlcolor{black}\href{https://doi.org/10.1021/acsami.1c22192}{\detokenize{10.1021/acsami.1c22192}}}.

\bibitem[{Al Zoubi} \em{et~al.}(2022){Al Zoubi}, Nashrah, {Kurnia Putri},
  Allaf, Assfour, and Ko]{AlZoubi2022a}
{Al Zoubi}, W.; Nashrah, N.; {Kurnia Putri}, R.A.; Allaf, A.W.; Assfour, B.;
  Ko, Y.G.
\newblock Strong dual-metal-support interactions induced by low-temperature
  plasma phenomenon.
\newblock {\em Materials Today Nano} {\bf 2022}, p. 100213.
\newblock
  doi:{\changeurlcolor{black}\href{https://doi.org/10.1016/j.mtnano.2022.100213}{\detokenize{10.1016/j.mtnano.2022.100213}}}.

\bibitem[Chen \em{et~al.}(2022)Chen, Chen, Yang, Wen, Yi, Zhou, Dai, Zhang, Wu,
  and Wu]{Chen2022}
Chen, H.; Chen, Z.; Yang, H.; Wen, L.; Yi, Z.; Zhou, Z.; Dai, B.; Zhang, J.;
  Wu, X.; Wu, P.
\newblock Multi-mode surface plasmon resonance absorber based on dart-type
  single-layer graphene.
\newblock {\em {RSC} Advances} {\bf 2022}, {\em 12},~7821--7829.
\newblock
  doi:{\changeurlcolor{black}\href{https://doi.org/10.1039/d2ra00611a}{\detokenize{10.1039/d2ra00611a}}}.

\bibitem[Zheng \em{et~al.}(2022)Zheng, Luo, Yang, Yi, Zhang, Song, Yang, Liu,
  Wu, and Wu]{Zheng2022}
Zheng, Z.; Luo, Y.; Yang, H.; Yi, Z.; Zhang, J.; Song, Q.; Yang, W.; Liu, C.;
  Wu, X.; Wu, P.
\newblock Thermal tuning of terahertz metamaterial absorber properties based on
  $VO_2$.
\newblock {\em Physical Chemistry Chemical Physics} {\bf 2022}, {\em
  24},~8846--8853.
\newblock
  doi:{\changeurlcolor{black}\href{https://doi.org/10.1039/D2CP01070D}{\detokenize{10.1039/D2CP01070D}}}.

\bibitem[Li \em{et~al.}(2021)Li, Yang, Wang, Zhang, and Li]{Li2021}
Li, R.; Yang, X.; Wang, Y.; Zhang, J.; Li, J.
\newblock Graphitic encapsulation and electronic shielding of metal
  nanoparticles to achieve metal--carbon interfacial superlubricity.
\newblock {\em ACS Applied Materials \& Interfaces} {\bf 2021}, {\em
  13},~3397--3407.
\newblock
  doi:{\changeurlcolor{black}\href{https://doi.org/10.1021/acsami.0c18900}{\detokenize{10.1021/acsami.0c18900}}}.

\bibitem[Xiao \em{et~al.}(2017)Xiao, Youji, Feitai, Peng, and Ming]{Xiao2017}
Xiao, L.; Youji, L.; Feitai, C.; Peng, X.; Ming, L.
\newblock Facile synthesis of mesoporous titanium dioxide doped by Ag-coated
  graphene with enhanced visible-light photocatalytic performance for methylene
  blue degradation.
\newblock {\em {RSC} Advances} {\bf 2017}, {\em 7},~25314--25324.
\newblock
  doi:{\changeurlcolor{black}\href{https://doi.org/10.1039/c7ra02198d}{\detokenize{10.1039/c7ra02198d}}}.

\bibitem[Zhang \em{et~al.}(2015)Zhang, Cai, Long, and Wang]{zhang2015}
Zhang, Z.; Cai, R.; Long, F.; Wang, J.
\newblock Development and application of tetrabromobisphenol A imprinted
  electrochemical sensor based on graphene/carbon nanotubes three-dimensional
  nanocomposites modified carbon electrode.
\newblock {\em Talanta} {\bf 2015}, {\em 134},~435--442.
\newblock
  doi:{\changeurlcolor{black}\href{https://doi.org/10.1016/j.talanta.2014.11.040}{\detokenize{10.1016/j.talanta.2014.11.040}}}.

\bibitem[Wang \em{et~al.}(2004)Wang, Luk’yanchuk, Hong, Lin, and
  Chong]{Luk2004:PRB}
Wang, Z.; Luk’yanchuk, B.; Hong, M.; Lin, Y.; Chong, T.
\newblock Energy flow around a small particle investigated by classical Mie
  theory.
\newblock {\em Physical Review B} {\bf 2004}, {\em 70},~035418.
\newblock
  doi:{\changeurlcolor{black}\href{https://doi.org/10.1103/PhysRevB.70.035418}{\detokenize{10.1103/PhysRevB.70.035418}}}.

\bibitem[Bashevoy \em{et~al.}(2005)Bashevoy, Fedotov, and
  Zheludev]{Zheludev2005:OE}
Bashevoy, M.; Fedotov, V.; Zheludev, N.
\newblock Optical whirlpool on an absorbing metallic nanoparticle.
\newblock {\em Optics express} {\bf 2005}, {\em 13},~8372--8379.
\newblock
  doi:{\changeurlcolor{black}\href{https://doi.org/10.1364/OPEX.13.008372}{\detokenize{10.1364/OPEX.13.008372}}}.

\bibitem[Luk'yanchuk \em{et~al.}(2006)Luk'yanchuk, Tribel'ski{\u{i}}, and
  Ternovski{\u{i}}]{OptJ_2006}
Luk'yanchuk, B.; Tribel'ski{\u{i}}, M.; Ternovski{\u{i}}, V.
\newblock Light scattering at nanoparticles close to plasmon resonance
  frequencies.
\newblock {\em Journal of Optical Technology} {\bf 2006}, {\em 73},~371--377.
\newblock
  doi:{\changeurlcolor{black}\href{https://doi.org/10.1364/JOT.73.000371}{\detokenize{10.1364/JOT.73.000371}}}.

\bibitem[Tribelsky and Luk’yanchuk(2006)]{Tribelsky:2006jf}
Tribelsky, M.I.; Luk’yanchuk, B.S.
\newblock Anomalous light scattering by small particles.
\newblock {\em Physical Review Letters} {\bf 2006}, {\em 97},~263902.
\newblock
  doi:{\changeurlcolor{black}\href{https://doi.org/10.1103/PhysRevLett.97.263902}{\detokenize{10.1103/PhysRevLett.97.263902}}}.

\bibitem[Novitsky and Barkovsky(2009)]{Novitsky2009}
Novitsky, A.V.; Barkovsky, L.M.
\newblock Poynting singularities in optical dynamic systems.
\newblock {\em Physical Review A} {\bf 2009}, {\em 79},~033821.
\newblock
  doi:{\changeurlcolor{black}\href{https://doi.org/10.1103/physreva.79.033821}{\detokenize{10.1103/physreva.79.033821}}}.

\bibitem[Angelsky \em{et~al.}(2021)Angelsky, Bekshaev, Hanson, Mokhun,
  Vasnetsov, and Wang]{Angelsky2021}
Angelsky, O.V.; Bekshaev, A.Y.; Hanson, S.G.; Mokhun, I.I.; Vasnetsov, M.V.;
  Wang, W.
\newblock Editorial: Singular and Correlation Optics.
\newblock {\em Frontiers in Physics} {\bf 2021}, {\em 9}.
\newblock
  doi:{\changeurlcolor{black}\href{https://doi.org/10.3389/fphy.2021.651964}{\detokenize{10.3389/fphy.2021.651964}}}.

\bibitem[Yue \em{et~al.}(2019)Yue, Yan, Monks, Dhama, Jiang, Minin, Minin, and
  Wang]{Yue2019}
Yue, L.; Yan, B.; Monks, J.N.; Dhama, R.; Jiang, C.; Minin, O.V.; Minin, I.V.;
  Wang, Z.
\newblock Full three-dimensional Poynting vector flow analysis of great
  field-intensity enhancement in specifically sized spherical-particles.
\newblock {\em Scientific Reports} {\bf 2019}, {\em 9},~20224.
\newblock
  doi:{\changeurlcolor{black}\href{https://doi.org/10.1038/s41598-019-56761-9}{\detokenize{10.1038/s41598-019-56761-9}}}.

\bibitem[Gao \em{et~al.}(2014)Gao, Novitsky, Zhang, Cheong, Gao, Lim,
  Luk{\textquotesingle}yanchuk, and Qiu]{Gao2014}
Gao, D.; Novitsky, A.; Zhang, T.; Cheong, F.C.; Gao, L.; Lim, C.T.;
  Luk{\textquotesingle}yanchuk, B.; Qiu, C.W.
\newblock Unveiling the correlation between non-diffracting tractor beam and
  its singularity in Poynting vector.
\newblock {\em Laser {\&} Photonics Reviews} {\bf 2014}, {\em 9},~75--82.
\newblock
  doi:{\changeurlcolor{black}\href{https://doi.org/10.1002/lpor.201400071}{\detokenize{10.1002/lpor.201400071}}}.

\bibitem[Dennis \em{et~al.}(2009)Dennis, O'Holleran, and Padgett]{Dennis2009}
Dennis, M.R.; O'Holleran, K.; Padgett, M.J.
\newblock Chapter 5 Singular Optics: Optical Vortices and Polarization
  Singularities; Elsevier,  2009; Vol.~53, {\em Progress in Optics}, pp.
  293--363.
\newblock
  doi:{\changeurlcolor{black}\href{https://doi.org/10.1016/S0079-6638(08)00205-9}{\detokenize{10.1016/S0079-6638(08)00205-9}}}.

\bibitem[Mokhun \em{et~al.}(2007)Mokhun, Khrobatin, Mokhun, and
  Viktorovskaya]{Mokhun2007}
Mokhun, I.; Khrobatin, R.; Mokhun, A.; Viktorovskaya, J.
\newblock The behavior of the Poynting vector in the area of elementary
  polarization singularities.
\newblock {\em Optica Applicata} {\bf 2007}, {\em 37},~261--277.

\bibitem[Bohren and Huffman(1998)]{Bohren::1998}
Bohren, C.F.; Huffman, D.R.
\newblock {\em Absorption and Scattering of Light by Small Particles};
  WILEY-VCH Verlag,  1998.

\bibitem[Landau \em{et~al.}(2013)Landau, Bell, Kearsley, Pitaevskii, Lifshitz,
  and Sykes]{landau2013electrodynamics}
Landau, L.D.; Bell, J.; Kearsley, M.; Pitaevskii, L.; Lifshitz, E.; Sykes, J.
\newblock {\em Electrodynamics of continuous media}; Vol.~8, Elsevier,  2013.

\bibitem[Polyanskiy()]{Polyanskiy}
Polyanskiy, M.
\newblock {\em Refractive index database}.
\newblock http://refractiveindex.info/.

\bibitem[Bohren(1983)]{bohren1983can}
Bohren, C.F.
\newblock How can a particle absorb more than the light incident on it?
\newblock {\em American Journal of Physics} {\bf 1983}, {\em 51},~323--327.

\bibitem[Tribelsky and Miroshnichenko(2022)]{tribelsky2022resonant}
Tribelsky, M.I.; Miroshnichenko, A.E.
\newblock Resonant scattering of electromagnetic waves by small metal
  particles: a new insight into the old problem.
\newblock {\em Physics-Uspekhi} {\bf 2022}, {\em 65},~40--61.
\newblock
  doi:{\changeurlcolor{black}\href{https://doi.org/10.3367/UFNe.2021.01.038924}{\detokenize{10.3367/UFNe.2021.01.038924}}}.

\end{thebibliography}

\end{document}